\documentstyle[editedvolume,epsfig,numreferences]{crckapb}

\begin{opening}
\title{Two-dimensional disk dynamos with vertical outflows into a halo}
\author{B. von Rekowski}
\author{W. Dobler}
\author{A. Shukurov}
\institute{Department of Mathematics, University of Newcastle\\
           Merz Court, Newcastle upon Tyne NE1 7RU, UK}
\author{A. Brandenburg}
\institute{Nordita, Blegdamsvej 17, DK-2100 Copenhagen \O, Denmark}
\end{opening}

\begin{document}

\begin{abstract}
We study the effects of vertical outflows on mean-field dynamos in
disks. These outflows could be due to thermal winds or magnetic buoyancy.
We analyse numerical solutions of the nonlinear mean-field dynamo equations
using a two-dimensional finite-difference model.
Contrary to expectations, a modest vertical velocity can enhance
dynamo action.
This can lead to super-exponential growth of the magnetic field and
to higher magnetic energies at saturation in the nonlinear regime.
\end{abstract}

\section{Introduction}
Large scale magnetic fields are often
considered to be an important factor in accretion disks
(including generation of turbulence), crucial in launching winds or jets.
The origin of such large scale magnetic fields in accretion
disks is still unclear: they may be advected from the surrounding
medium or be generated by a dynamo inside the disk. Advection
appears unlikely as turbulence leads to enhanced viscosity and
magnetic diffusivity, so that the two are of the same order, i.e.
the magnetic Prandtl number is of order unity \cite{PFL76}. In this
case the turbulent magnetic diffusivity can compensate the
dragging of the field by viscously induced accretion flow
\cite{vBa89,LPP94,HPB96}. Turbulent dynamo action is a
plausible mechanism for producing large scale magnetic fields in
accretion disks \cite{Pud81,SL88}. Dynamo magnetic fields can
launch winds from accretion disks \cite{CPA98,Cam99,BDSvR00}.
However, the wind can also affect the dynamo. In particular, a
dynamo enhancement by winds was suggested earlier for galactic
dynamos \cite{Bra93}.
In the same context the effects of shear in the vertical velocity
on the dynamo was considered \cite{EGRW95}.

We assume here that the magnetic fields are generated by a dynamo acting
in a relatively thin accretion disk. We show how vertical flow can enhance
the dynamo, allowing for a larger growth rate and leading to super-exponential
growth of the magnetic field and to enhanced saturation levels of
magnetic energy.

The vertical velocities can have several origins.
For example, they can be due to a thermally driven wind emanating
from the disk or magnetic buoyancy in the disk.
Note that here we invoke magnetic buoyancy as a driver of vertical
outflows; it can itself contribute to the dynamo effect \cite{MSS99},
but such effects are not taken into account here.

\section{The model}
The equation which we solve is the mean-field induction equation
which we evolve in terms of the vector potential
${\bf A}$, where ${\bf B} = {\bf \nabla} \times {\bf A}$,
\begin{equation}
{\partial {\bf A} \over \partial t} = {\bf V} \times \left({\bf \nabla}
\times {\bf A}\right)
+ \alpha {\bf \nabla} \times {\bf A}
- \eta \mu_0 {\bf j},
\label{eqA}
\end{equation}
where ${\bf j} = {\bf \nabla} \times {\bf B} / \mu_0$ is the
current density, $\mu_0$ the magnetic permeability, $\eta$ the
turbulent magnetic diffusivity, $\alpha$ the $\alpha$-effect and
${\bf V}$ the mean velocity. We neglect the radial component of the mean
velocity. Assuming the $\alpha$- and $\eta$-tensors to be
isotropic, we can consider scalar quantities. We do not make the
thin-disk approximation but solve the general equations. We adopt
cylindrical coordinates $(r, \varphi, z)$ and restrict ourselves
to axisymmetric solutions. Equation~(\ref{eqA}) is solved using a sixth order
finite-difference scheme in space and a third order Runge-Kutta time advance scheme.

Our computational domain contains a disk embedded in its surrounding halo.
We take a disk aspect ratio of $h_{\rm disk}/R_{\rm disk} = 0.1$ and a halo
with $h_{\rm halo}/h_{\rm disk} \approx 6$.

As an initial condition for the magnetic field we choose a purely
poloidal field in the disk of either dipolar or quadrupolar
symmetry.

On the boundaries of the computational domain we impose pseudo-vacuum
conditions. However, since the boundaries are in the halo far away
from the disk, the choice of boundary conditions is not crucial.

The $\alpha$-coefficient $\alpha(r,z)$ is antisymmetric about
the disk midplane and vanishes outside the disk. We adopt
\begin{equation}
\alpha(r,z) =
\left\{ \begin{array}{cl}
        \alpha_0(r) \sin \left(\pi {z \over h}\right) \xi_\alpha(r) &
    \mbox{for $|z| \leq h$}, \\
    0 &
    \mbox{for $|z| > h$}.
    \end{array} \right.
\label{alp}
\end{equation}
The $\xi_\alpha$-profile cuts off the $\alpha$-effect at the outer
radius of the disk as well as at an inner radius, where the
rotational shear is very strong.

As appropriate for accretion disks, we adopt a softened Keplerian angular
velocity profile in $r$ in the disk as well as the halo,
\begin{equation}
\Omega(r) = \sqrt{GM \over r^3}
\left[ 1 + \left( {r_0 \over r} \right)^n \right]^{-{n+1 \over 2n}},
\end{equation}
where $G$ is Newton's gravitational constant, $M$ is the mass of the central object,
$r_0 = 0.05$ is the softening radius, and $n=5$. At $r = 0$, $\Omega$ vanishes.

The turbulent magnetic diffusivity is given by
\begin{equation}
\eta(r,z) = \eta_{\rm halo} + (\eta_{\rm disk} - \eta_{\rm halo}) \xi(r,z).
\end{equation}
The profile $\xi(r,z)$ defines the disk: $\xi$ is equal to unity inside the disk
and vanishes in the halo but with a smooth transition between.
We carried out computations for two cases:
homogeneous conductivity, $\eta_{\rm halo}/\eta_{\rm disk} = 1$, and
low conductivity in the halo, $\eta_{\rm halo}/\eta_{\rm disk} = 20$.

The profile for $V_z$ is linear in $z$,
$V_z(z) = V_{z0} z / h$,
where $V_{z0}$ is a characteristic vertical velocity.

Our dynamo problem is controlled by the three magnetic Reynolds numbers
related to the $\alpha$-effect, the differential rotation and the vertical velocity,
\begin{equation}
R_\alpha = h \alpha_0 / \eta_{\rm disk}, \quad
R_\omega = h^2 S / \eta_{\rm disk}, \quad
R_v = h V_{z0} / \eta_{\rm disk},
\label{R123}
\end{equation}
where $S(r) \equiv r d\Omega / dr \approx -3/2 \ \Omega(r)$ is the rotational shear.
The dynamo number is defined as ${\cal D} \equiv R_\alpha R_\omega$.
Since the dynamo number is approximately constant,
$|{\cal D}| \simeq \alpha_{\rm SS}^{-2} \simeq 10^2 - 10^4$ in thin accretion disks
\cite{Pud81},
we assume ${\cal D}$ to be constant by setting
$\alpha_0(r) = {\cal D} \eta_{\rm disk}^2 / S(r) h^3$.

Because of the strong differential rotation
we assume that the magnetic field in the disk is generated by a
standard $\alpha\omega$-dynamo.
The two control parameters are then the dynamo number $\cal D$ and
the vertical magnetic Reynolds number $R_v$.
The value of $R_v$ obviously depends on the origin of the vertical outflow.
With magnetic buoyancy, a rough estimate can be done assuming that
the vertical velocities are comparable to the Alfv\'en speed. Estimating
the Alfv\'en speed from magnetic equipartition, one gets a vertical
magnetic Reynolds number $R_v$ of order unity.

\section{Linear results for $R_v = 0$}
\begin{figure}[t]
\centering
\epsfig{figure=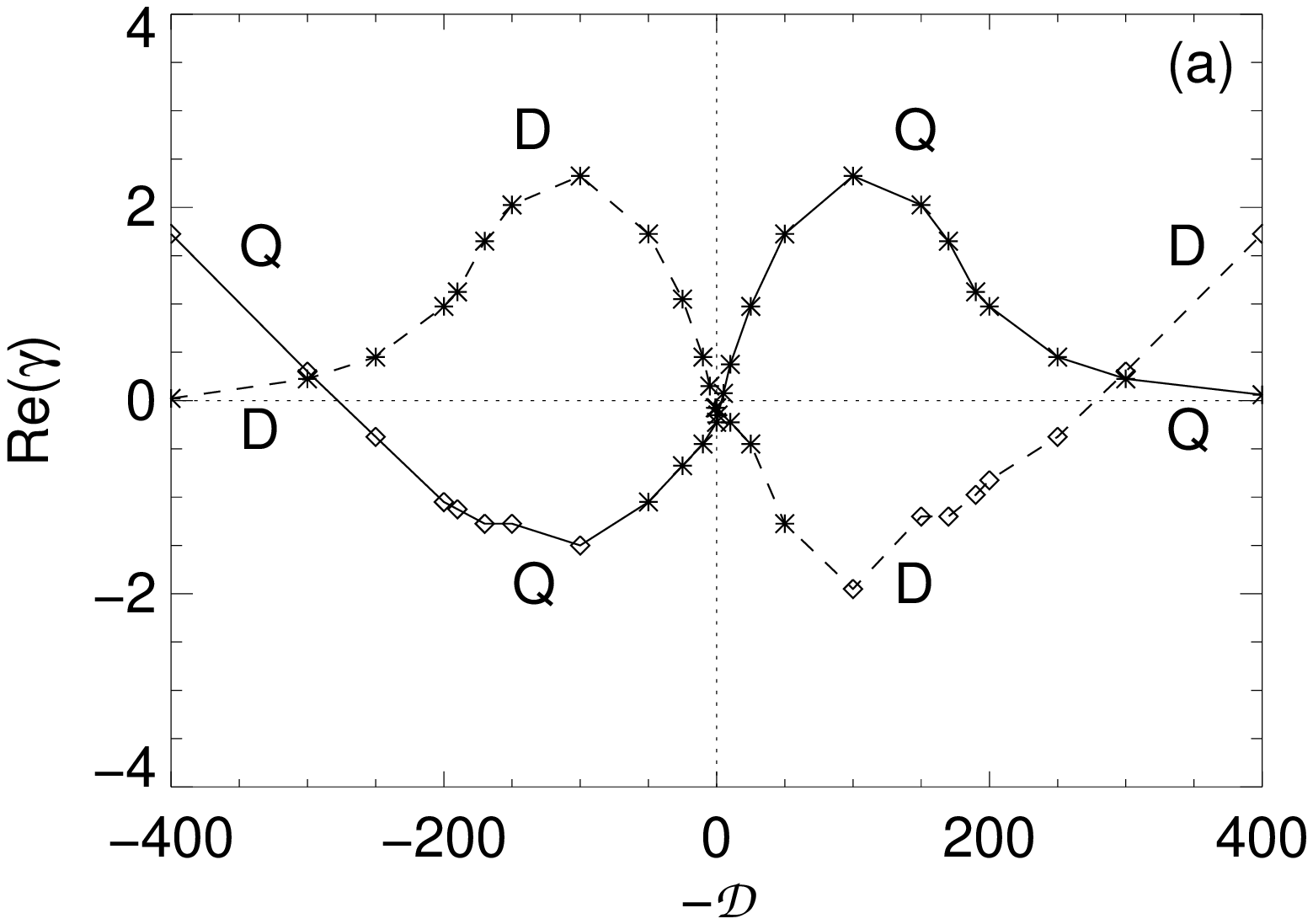,height=5cm,width=6.0cm}
\epsfig{figure=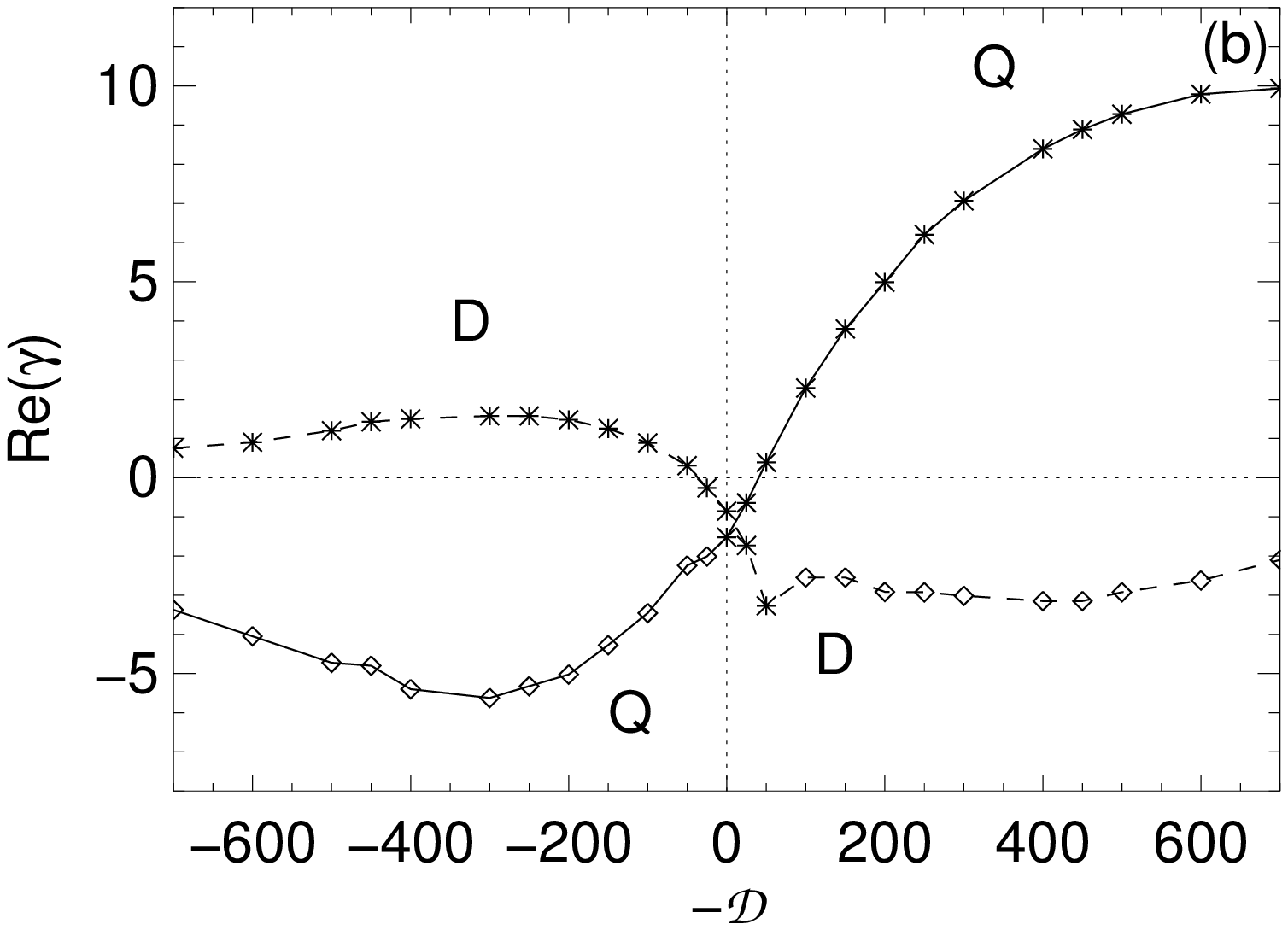,height=5cm,width=6.0cm}
\caption{ Real part of the growth rate of the magnetic field as a
function of $-{\cal D}$ with $R_v = 0$. Asterisks denote
non-oscillatory, diamonds oscillatory solutions. Solid lines
denote quadrupolar (Q), dashed lines dipolar (D) modes. (a) is
with homogeneous conductivity, (b) for low-conducting halo. Note
the symmetry of (a) with respect to ${\cal D} = 0$.}
\label{fig1}
\end{figure}

For $\eta_{\mathrm{halo}} = \eta_{\mathrm{disk}}$,
the growth rates of dipolar and quadrupolar modes should be exchanged when ${\cal D}$
reverses sign, i.e. the graph of $Re(\gamma)$ as a function of ${\cal D}$ should be
symmetric with respect to the vertical axis ${\cal D} = 0$ with quadrupolar and dipolar
modes exchanging their r\^oles \cite{Pro77}. Our numerical simulation shows
this symmetry with a good precision (Fig.~\ref{fig1}a).
This is because the ratio between $h_{\rm halo}$ and $h_{\rm disk}$ is large enough,
about 6.

For ${\cal D} < 0$ ($R_\alpha > 0$) the first leading growing mode is quadrupolar
whereas for ${\cal D} > 0$ ($R_\alpha < 0$) it is dipolar; both are non-oscillatory.
We thus study the effect of vertical velocities
for ${\cal D} < 0$ in quadrupolar and for ${\cal D} > 0$ in dipolar symmetry
at dynamo numbers up to $\pm 300$.

For $\eta = \rm{const.}$, the critical dynamo number is $|{\cal D}_{\rm crit}|
\simeq 5$. The dominant mode changes its symmetry and
becomes oscillatory at $|{\cal D}| \simeq 300$.

The qualitative behaviour in the case of low conductivity in the halo is quite
similar to that of homogeneous conductivity.
Also the quadrupolar non-oscillatory modes first become dominant for $R_\alpha > 0$ and
the dipolar non-oscillatory modes for $R_\alpha < 0$ (compare Figs.~\ref{fig1}a and b).

But there are quantitative differences. The value of $|{\cal D}|$,
where the symmetry of the leading mode changes, is larger than $700$.
Also, the diagram in Fig.~\ref{fig1}b is not symmetric.
The critical dynamo number is ${\cal D} \simeq -50 \ (50)$ for
quadrupolar (dipolar) modes.

\section{Linear behaviour with vertical velocities}
\begin{figure}[t]
\centering
\epsfig{figure=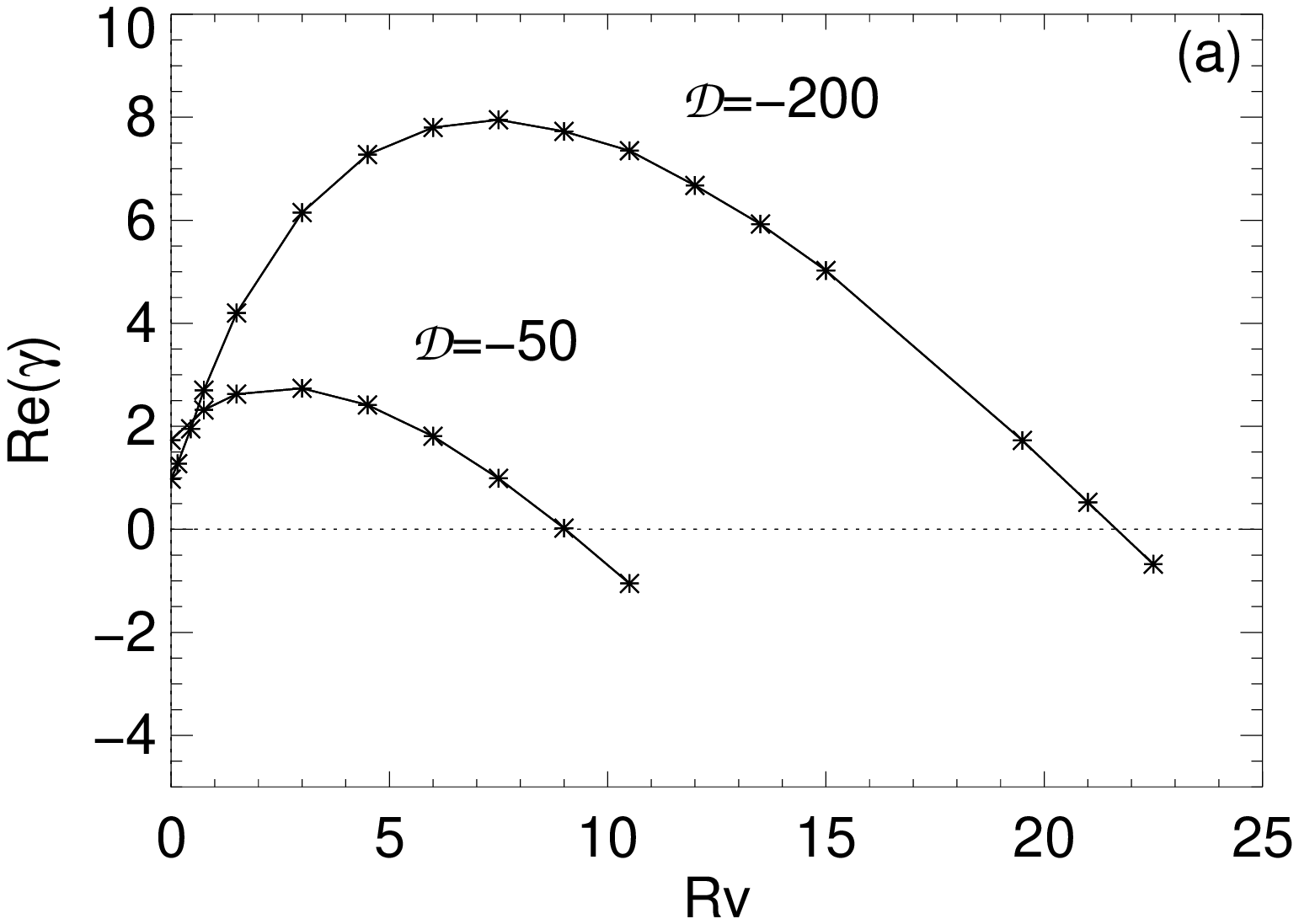,height=5cm,width=6.0cm}
\epsfig{figure=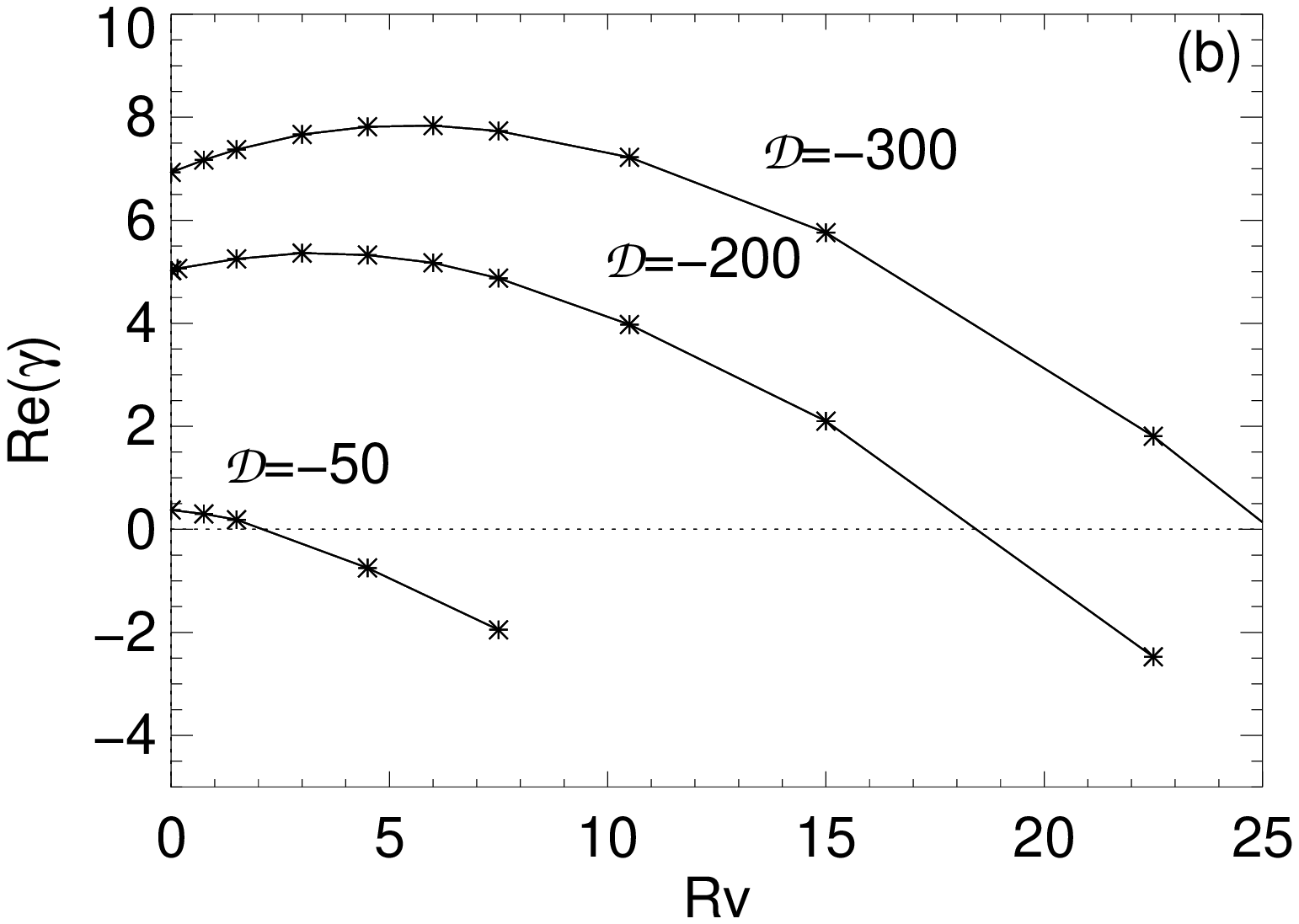,height=5cm,width=6.0cm}
\caption{Real part of the growth rate of the magnetic field as a
function of $R_v$ for non-oscillatory quadrupolar modes. (a)
$\eta_{\rm halo} / \eta_{\rm disk} = 1$, (b) $\eta_{\rm halo} /
\eta_{\rm disk} = 20$.} \label{fig2}
\end{figure}

As discussed in section 3, we only consider quadrupolar symmetry
for ${\cal D} < 0$.
As shown in Fig.~\ref{fig2}, the growth rate of the magnetic field ${\rm Re} (\gamma)$
is a non-monotonous function of $R_v$, for $\eta_{\rm halo} / \eta_{\rm disk}$
equal to both $1$ and $20$. A maximum in ${\rm Re} (\gamma)$ occurs, however,
only for $|{\cal D}|$ large enough, and the smaller the
magnetic diffusivity of the halo, the smaller is the required $|{\cal D}|$.
The larger $|{\cal D}|$, the more pronounced is
the maximum. The maximum growth rate occurs for $R_v$ of the order
of $1$ to $10$.

\begin{figure}[t]
\centering \epsfig{figure=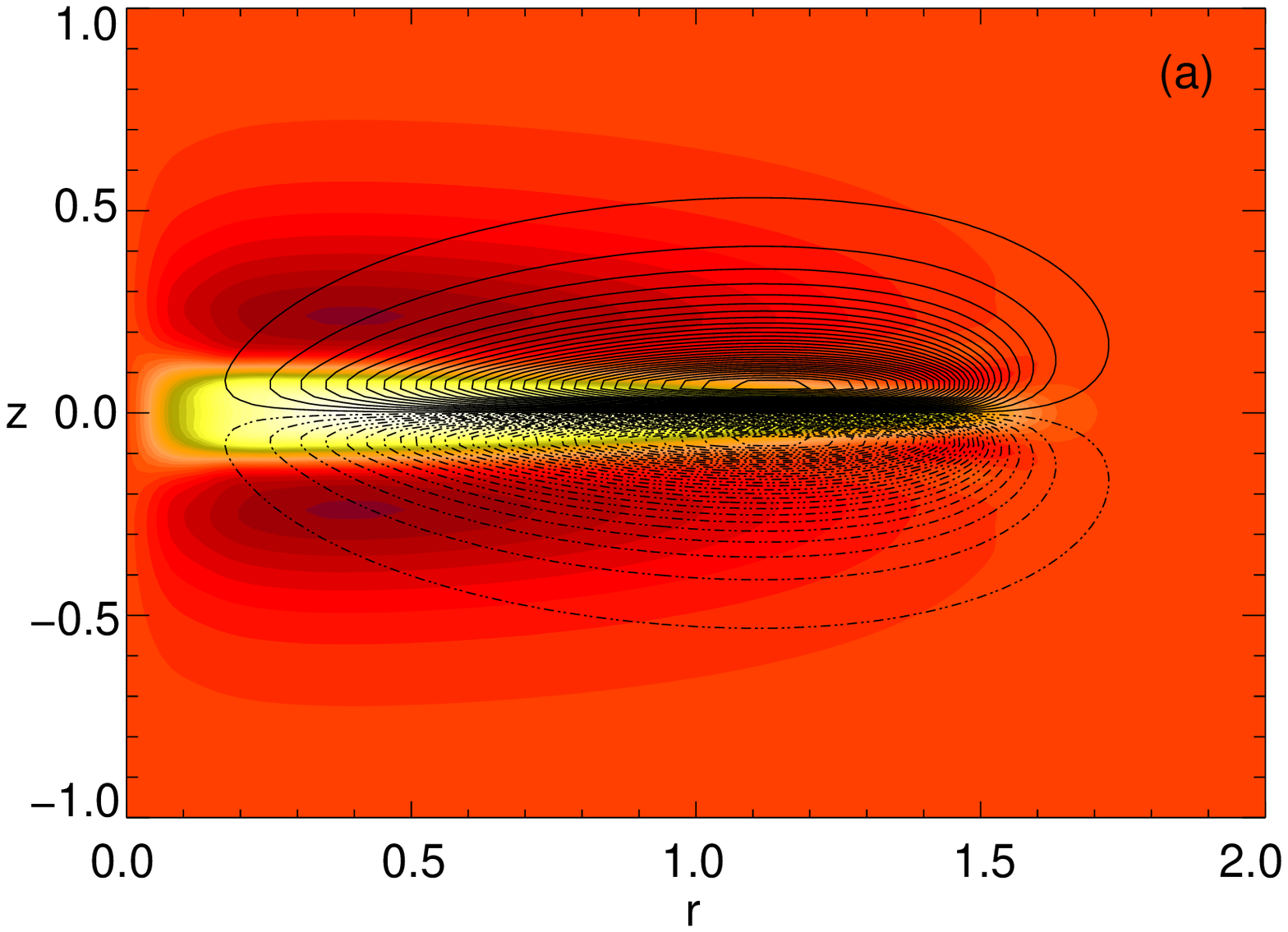,height=5cm,width=5cm}
\epsfig{figure=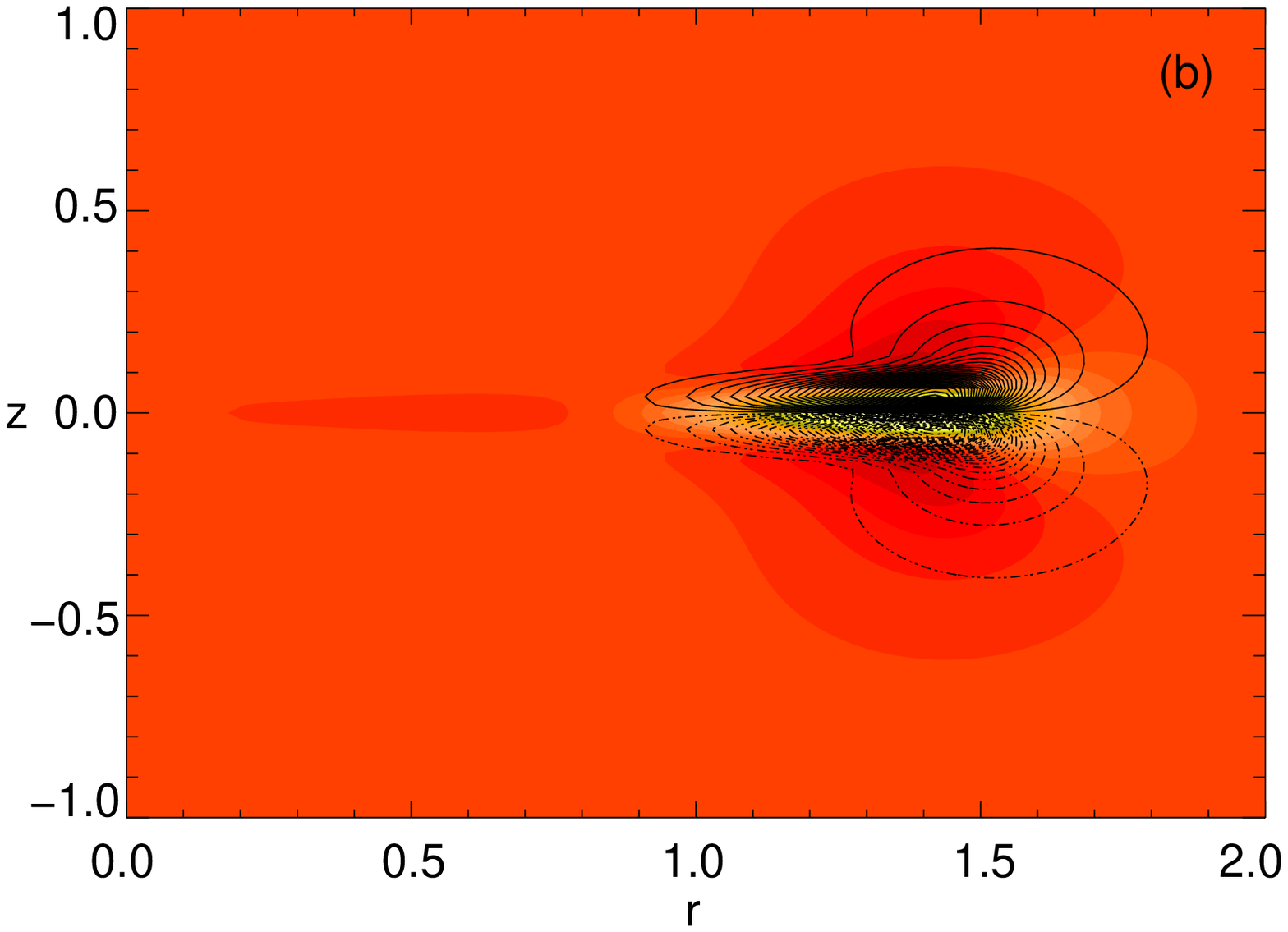,height=5cm,width=5cm}
\epsfig{figure=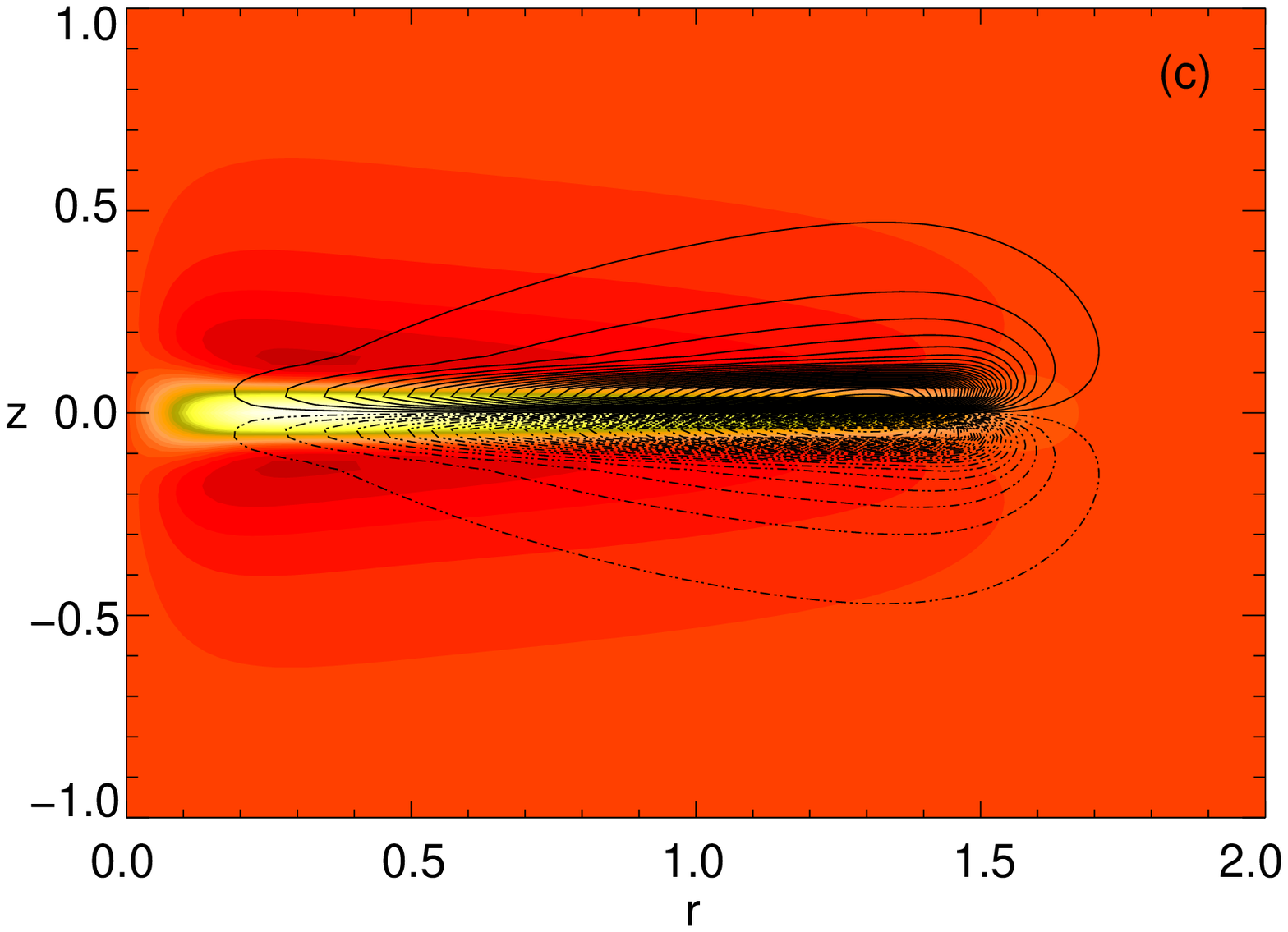,height=5cm,width=5cm}
\epsfig{figure=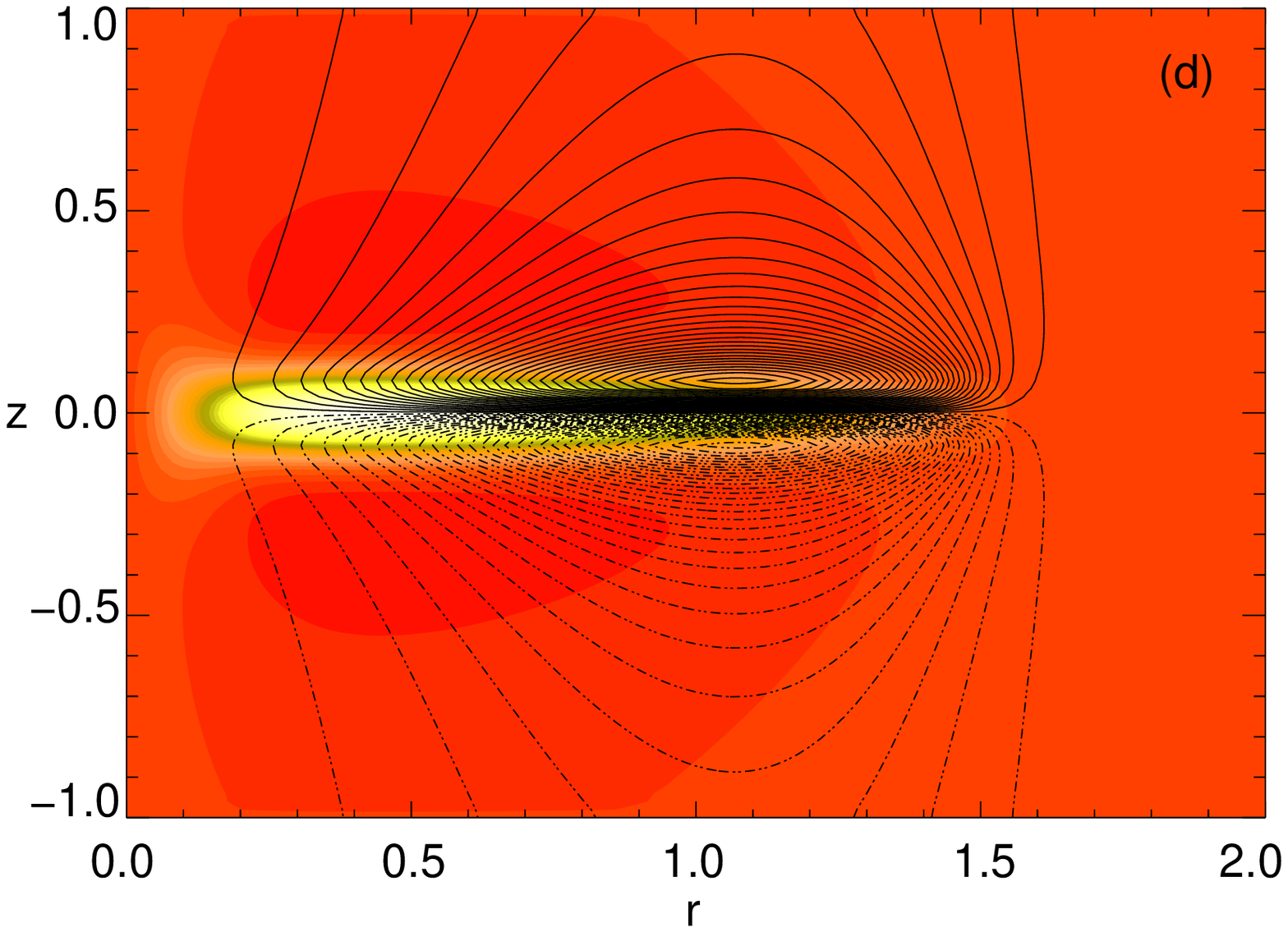,height=5cm,width=5cm} \caption{
The effect of dynamo number and vertical velocity on the magnetic
field configuration, for the linear model with $\eta_{\rm halo} =
\eta_{\rm disk}$. Shown are the poloidal field lines (solid is
clockwise, dotted is counter-clockwise) and the toroidal field
(grey scales: bright is in positive, dark in negative azimuthal
direction). (a): ${\cal D} =  -50$, $R_v = 0$; (b): ${\cal D} =
-200$, $R_v = 0$; (c): ${\cal D} = -200$, $R_v = 0.75$; (d):
${\cal D} = -200$, $R_v = 7.5$ (where Re$(\gamma)$ is maximum).
Comparing (a) and (b) shows the effect of increasing ${\cal D}$
while $R_v=0$. Panels (b) to (d) follow the upper curve in
Fig.~\ref{fig2}a. All modes are quadrupolar and non-oscillatory.}
\label{fig4}
\end{figure}

In Fig.~\ref{fig4} we show the effect of the dynamo number and the
vertical velocity on the magnetic field configuration for
$\eta_{\rm halo} = \eta_{\rm disk}$. Increasing
$|{\cal D}|$ from $50$ to $200$ in the absence of any vertical
velocity results in the magnetic field becoming concentrated at
larger radii in the disk (Fig.~\ref{fig4}a and b).

As can be seen in Fig.~\ref{fig4}b, c and d, a vertical velocity
with $R_v < 7.5$ leads to a wider vertical distribution of magnetic
field which reduces magnetic diffusion and enhances the dynamo action.
At still larger $R_v$,
the wind aligns the poloidal field lines with the lines of
constant rotation, and the $\Omega$-effect is slowly switched off
leading to a decrease in the growth rate, which eventually becomes
negative. The mode structures at a maximum of Re($\gamma$), reached at
$R_v = 7.5$ for ${\cal D} = -200$) and $R_v = 3$ for ${\cal D} = -50$,
are very similar to each other.

The growth rate for dipolar non-oscillatory modes (${\cal D} > 0$)
has no maximum as $R_v$ increases; at $R_v$ of order $1$ or even less
the dynamo is switched off. The magnetic field is advected outwards
and aligned with the $\Omega$-contour lines in vertical direction
very quickly.

\section{Nonlinear behaviour with vertical velocities}
We consider nonlinear, saturated solutions in a model with
homogeneous conductivity, $\eta_{\rm halo} = \eta_{\rm
disk}$, negative dynamo number and quadrupolar symmetry (the
dominant symmetry for moderate $|{\cal D}|$ with $R_v = 0$). All
saturated magnetic fields are non-oscillatory, as the
corresponding linear modes.

\subsection{Magnetic buoyancy}
We parameterize the effect of magnetic buoyancy by assuming the
vertical velocity to be proportional to the maximum magnetic field
strength. Thus, the vertical magnetic Reynolds number becomes
time-dependent,
\begin{equation}
R_v = R_{v0} \max \limits_{\bf x} (B_r,B_\varphi) /
|B_0|, \quad R_{v0} \equiv V_{z0} h / \eta_{\rm disk},
\label{rel}
\end{equation}
where $B_0$ is a characteristic field strength.

As time increases, $|{\bf B}|$ grows and
therefore $R_v$ is increasing and the growth rate follows, e.g., the
upper curve in Fig.~\ref{fig2}a for ${\cal D} = -200$. Thus,
as long as $R_v$ is less than the position of the $\gamma$-maximum,
$\gamma$ increases with time, which results in a super-exponential
growth. After the maximum, at $R_v = 7.5$, $\gamma$ decreases and
eventually the magnetic field approaches its saturation level. This is
reached when $R_v = R_{v*}$, where $R_{v*}$ is the zero of $\gamma(R_v)$.
According to Eq.~(\ref{rel}), the saturation value of magnetic energy
will thus be $E_{\rm mag} \propto 1 / R_{v0}^2$.


\subsection{$\alpha$-quenching}
We consider the back-reaction of the magnetic field on the
$\alpha$-effect, introducing the nonlinearity
$1 / (1 + B^2 / B_{\rm eq}^2)$ as a factor in the $\alpha$-effect;
$B_{\rm eq}$ is the equipartition field.

\begin{figure}[t]
\centering \epsfig{figure=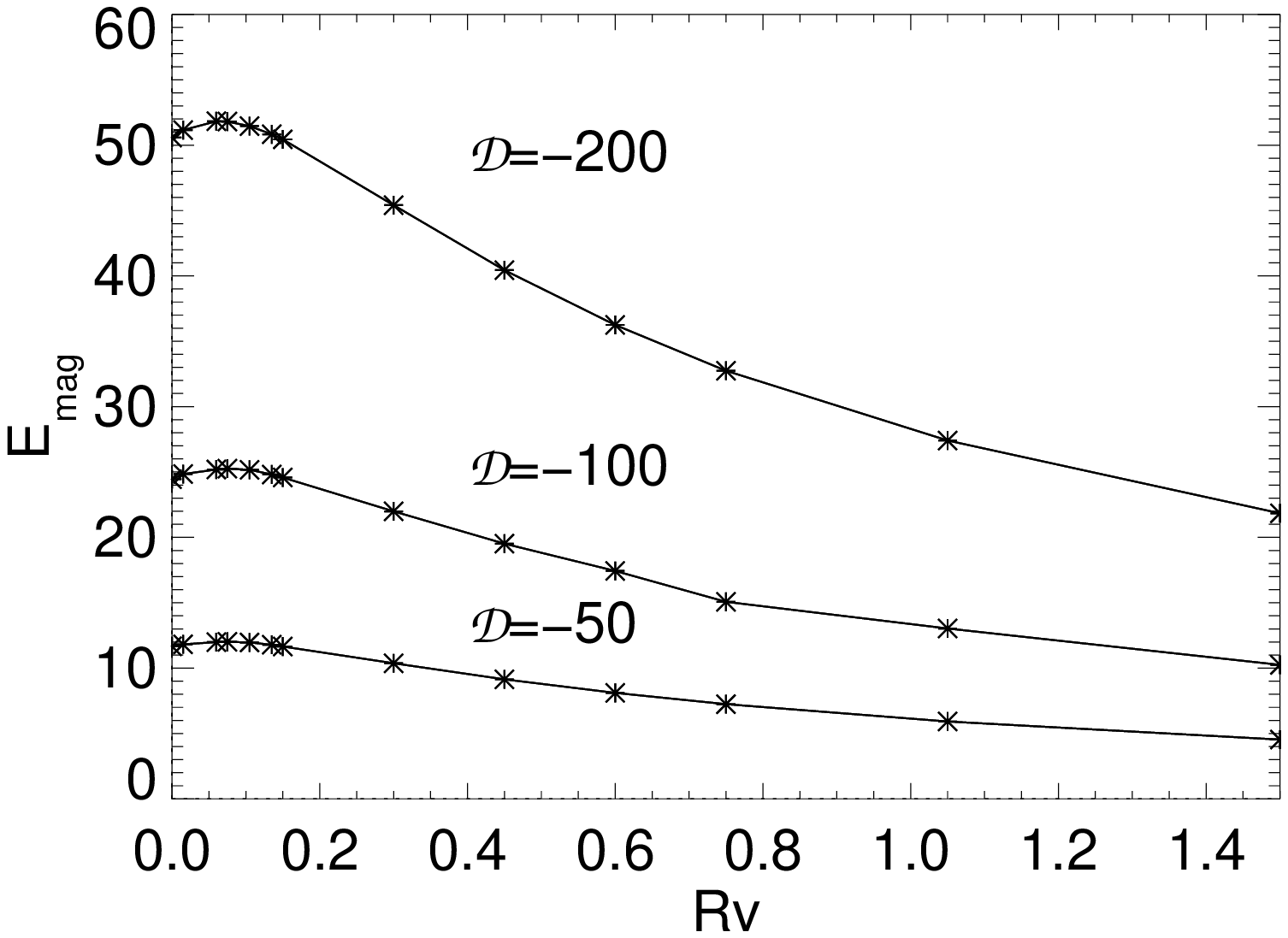,height=5cm,width=6.0cm}
\epsfig{figure=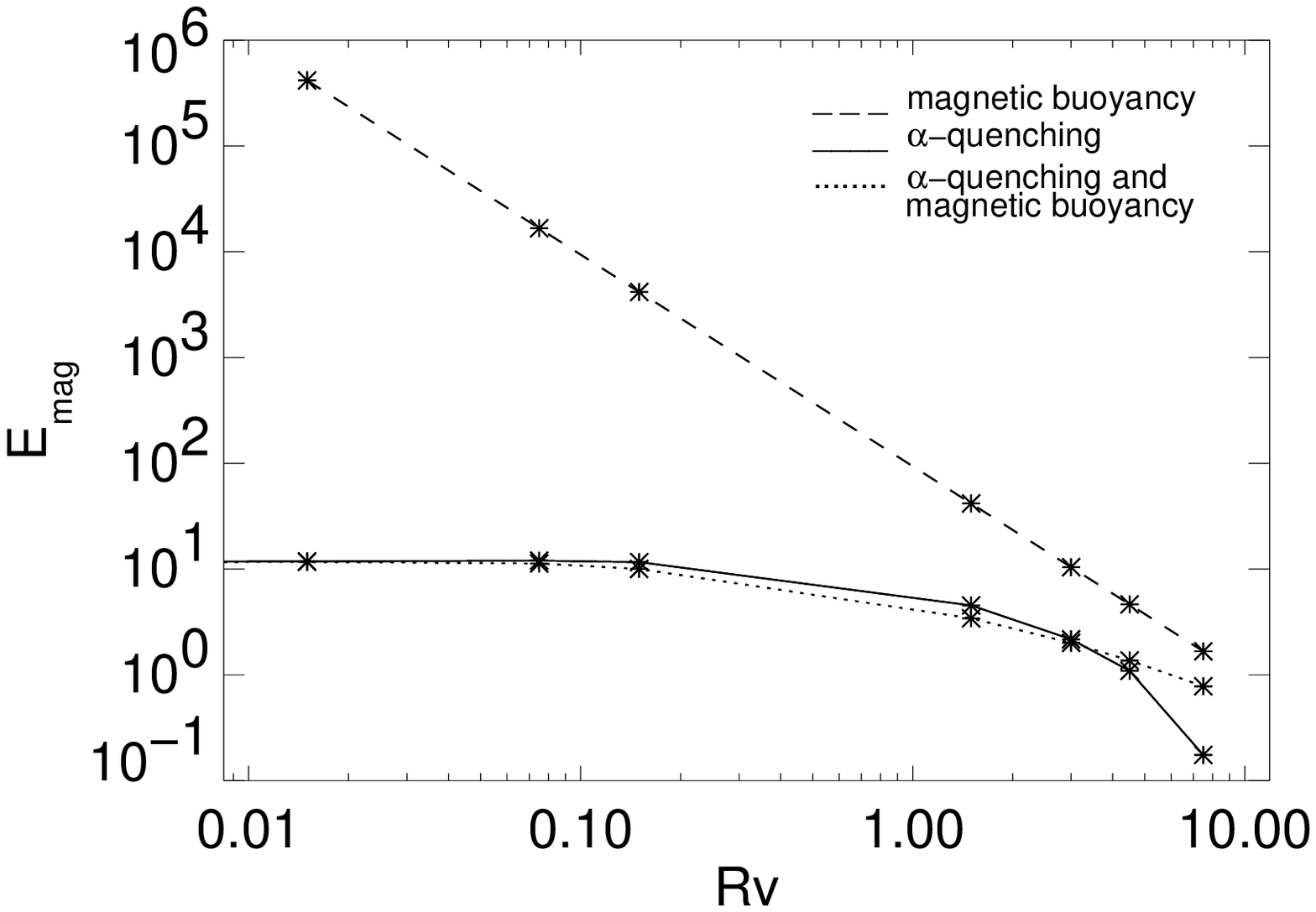,height=5cm,width=6.0cm}
\caption{Magnetic energy as a function of $R_v$ for the model with
$\alpha$-quenching (left) and for the models with
$\alpha$-quenching and/or magnetic buoyancy (right) for $\eta_{\rm
halo} = \eta_{\rm disk}$. All modes are quadrupolar and
non-oscillatory. The right panel shows the case ${\cal D} = -50$.}
\label{fig911}
\end{figure}

The magnetic energy has a maximum at a certain value of $R_v$
(Fig.~\ref{fig911}, left), but this maximum occurs at
$R_v \approx 0.1$, a value smaller than where a maximum of Re($\gamma$)
occurs in the linear regime, and it is also less pronounced. The
value of $R_v$ where the maximum occurs, is independent of ${\cal
D}$. The magnetic energy scales roughly with the dynamo number,
$E_{\rm mag} \propto {\cal D}$.

\begin{figure}[t]
\centering \epsfig{figure=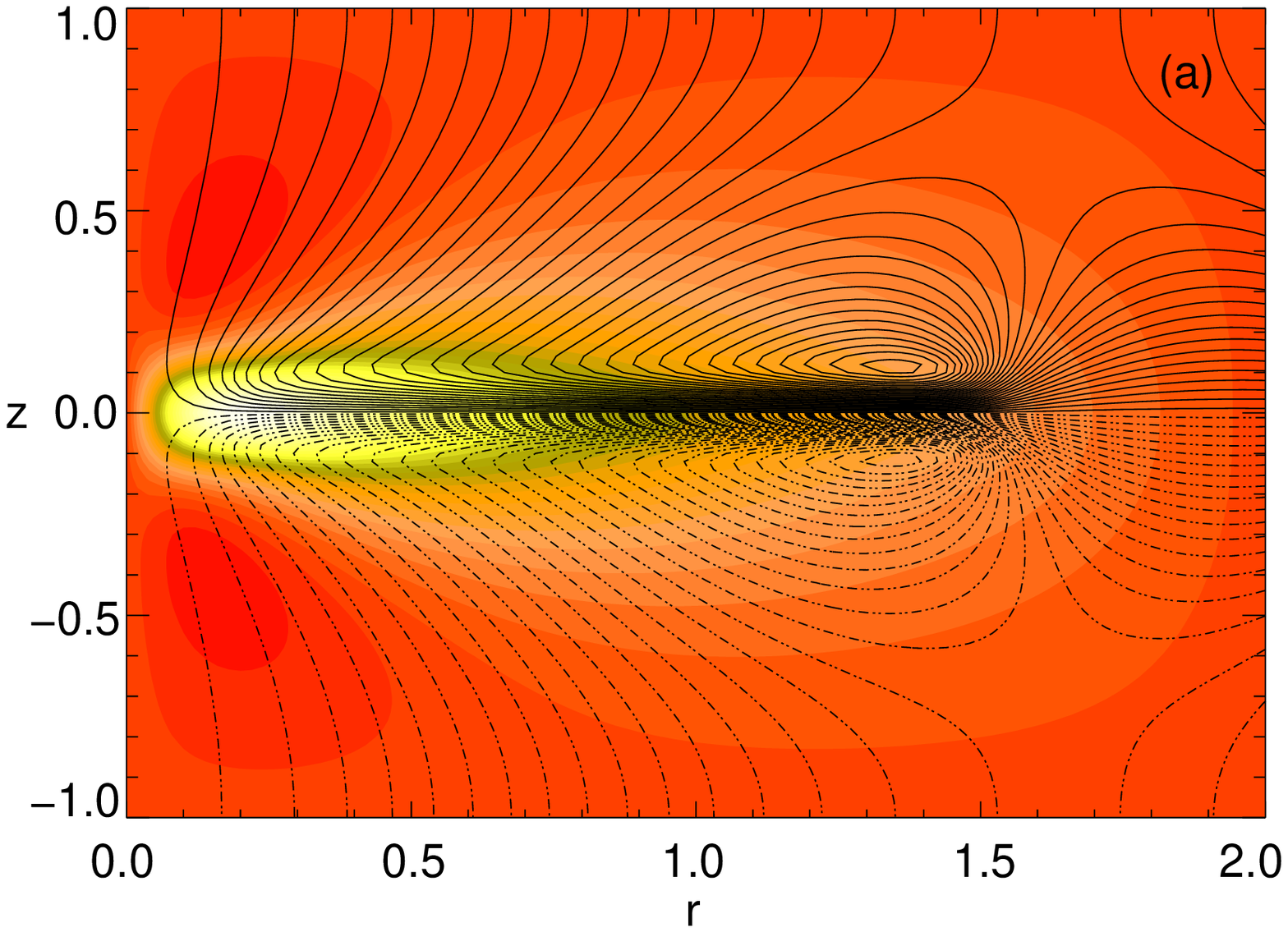,height=3.7cm,width=3.7cm}
\epsfig{figure=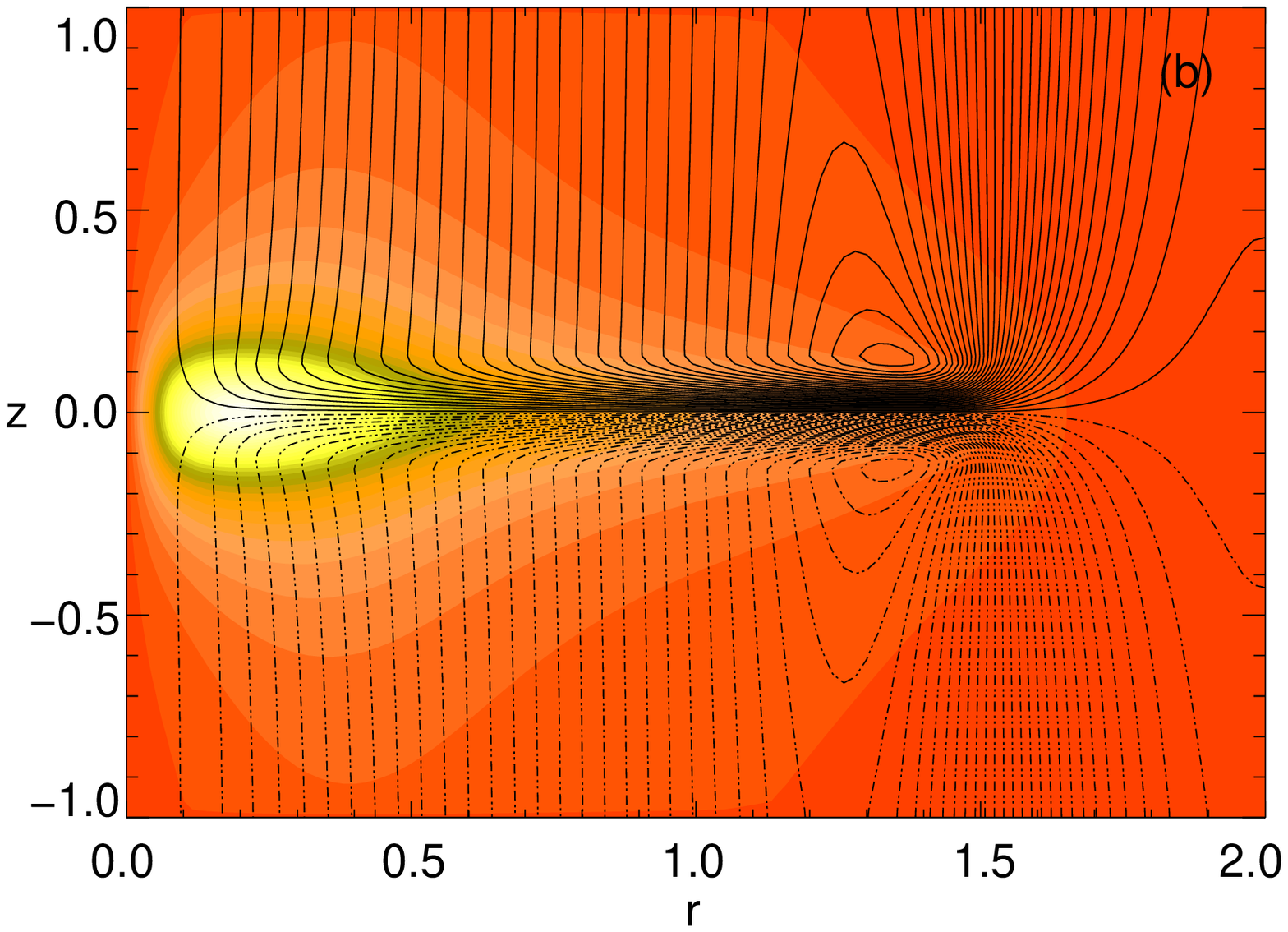,height=3.7cm,width=3.7cm}
\epsfig{figure=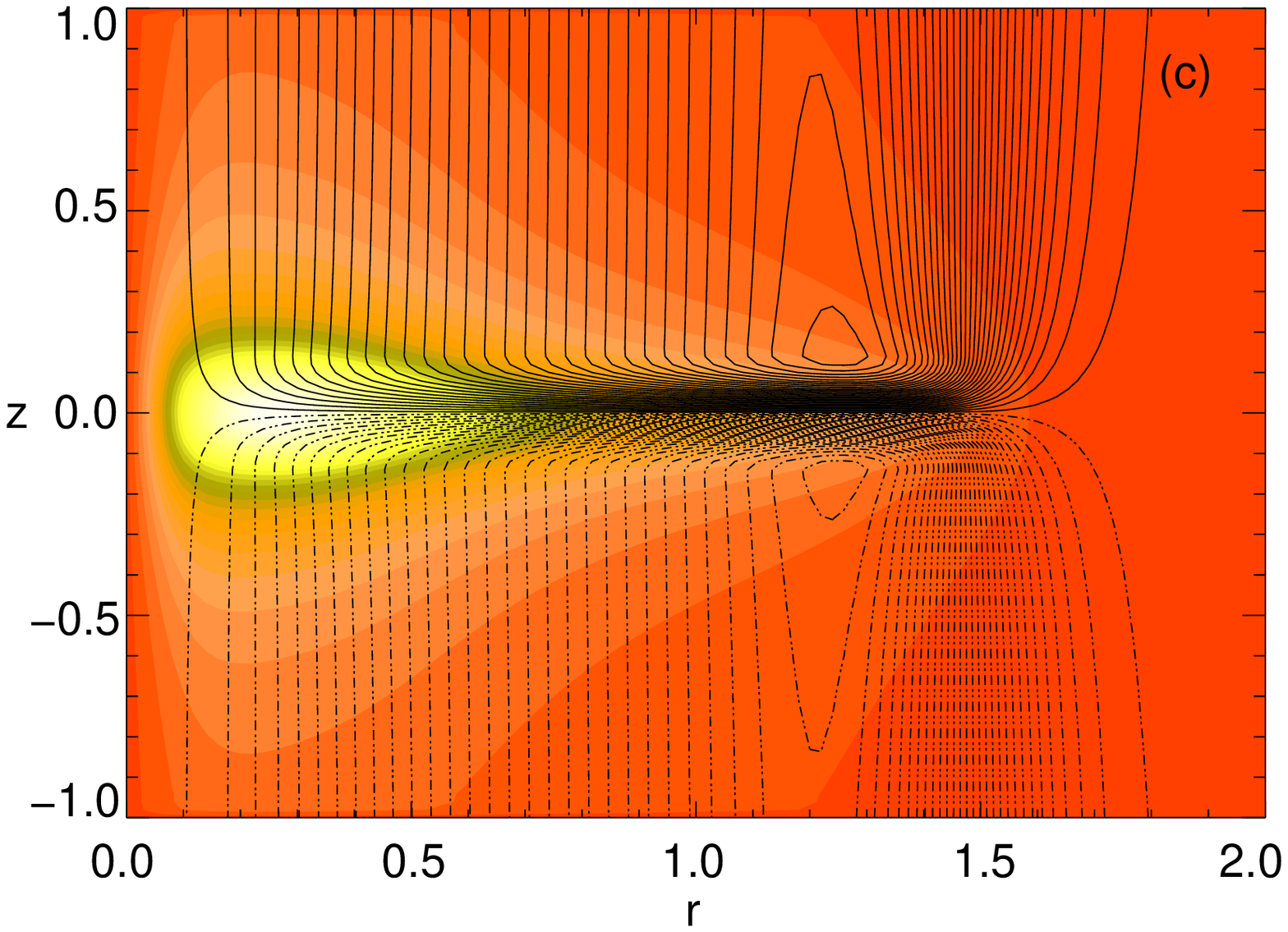,height=3.7cm,width=3.7cm} \caption{The
effect of vertical velocity on the magnetic field configuration
for the model with $\alpha$-quenching and $\eta_{\rm halo} =
\eta_{\rm disk}$. Shown are the poloidal field lines
and the toroidal field. (a): ${\cal D} = -50$, $R_v = 0$;
(b): ${\cal D} = -50$, $R_v = 3$; (c): ${\cal D} = -50$, $R_v = 7.5$.
Panels (a) to (c) follow the lower curve in Fig.~\ref{fig911}, left.
All modes are quadrupolar and non-oscillatory.} \label{fig10}
\end{figure}

The effect of vertical velocity on the magnetic field for the
model with $\alpha$-quenching is shown in Fig.~\ref{fig10}; the
structure changes only weakly with the dynamo number. With
increasing $R_v$ the poloidal field lines become vertical, i.e.
aligned with the lines of constant rotation.
Hence, shear has no effect, and the $\Omega$-effect is switched off,
as in the linear model.

\subsection{$\alpha$-quenching together with magnetic buoyancy}
When $\alpha$-quenching and magnetic buoyancy are combined,
the vertical magnetic Reynolds number depends on position and time
and takes the form $R_v = R_{v0} \sqrt{B_r^2 + B_\varphi^2} / |B_{\rm eq}|$.
Without $\alpha$-quenching, since $R_v$ is now nonlocal, the
temporal behaviour of the growth rates at short times does not
have to be super-exponential. Surprisingly, again we find
$E_{\rm mag} \propto 1 / R_{v0}^2$ (see Fig.~\ref{fig911}, right,
dashed line).

In the second panel of Fig.~\ref{fig911} magnetic energy is plotted against
$R_v$ for ${\cal D} = -50$ for the three nonlinear models.
$\alpha$-quenching appears to be the dominant nonlinearity
in the model considered.

\section*{Acknowledgements}
We acknowledge financial support from PPARC (Grant PPA/G/S/1997/ 00284) and
the Leverhulme Trust (Grant F/125/AL).
The use of the PPARC supported GRAND parallel computer is acknowledged.

\end{document}